# Comparing Spectrum Utilization using Fuzzy Logic System for Heterogeneous Wireless Networks via Cognitive Radio

R. Kaniezhil, Dr. C. Chandrasekar

**Abstract**— At present, lots of works focus on spectrum allocation of wireless networks. In this paper, we proposed a Cognitive based spectrum access by opportunistically approach of Heterogeneous Wireless networks based on Fuzzy Logic system. The Cognitive Radio is a technology where a network or a wireless system changes its environment parameters to communicate efficiently by avoiding the interference with the users. By applying FLS (Fuzzy Logic System), the available spectrum utilization is effectively utilized with the help of the three antecedents namely Spectrum utilization efficiency, Degree of mobility, Distance from primary user to the secondary users. The proposed work is compared with normal Spectrum Utilization method. Finally, Simulation results of the proposed work Fuzzy Logic System shows more efficient than the normal Spectrum utilization method.

**Index Terms**— CR, FLS, Interference, Spectrum access, Secondary users, Spectrum utilization.

—————— ◆ ——————

## 1 INTRODUCTION

In recent studies, the spectrum allocated by the traditional approach shows that the spectrum allocated to the primary user is under-utilized and the demand for accessing the limited spectrum is growing increasingly [13]. Spectrum is no longer sufficiently available, because it has been assigned to primary users that own the privileges to their assigned spectrum. However, it is not used efficiently most of the time. In order to use the spectrum in an opportunistic manner and to the increase spectrum availability, the unlicensed users can be allowed to utilize licensed bands of licensed users, without causing any interference with the assigned service.

The reason for allowing the unlicensed users to utilize licensed bands of licensed users if they would not cause any interference with the assigned service. This paradigm for wireless communication is known as opportunistic spectrum access and this is considered to be a feature of Cognitive Radio (CR). Cognitive radio is an emerging wireless communication paradigm in which either the Network or the wireless node itself intelligently adapts particular transmission or reception parameters by sensing the environment.

Dynamic spectrum access using CR is an emerging research topic. Cognitive radio techniques provide the capability to use or share the spectrum in an opportunistic manner.

With the growing number of wireless devices and increased spectrum occupancy, the unlicensed spectrum is getting scarce.

- R. Kaniezhil is currently pursuing Ph.D in Computer Science, Periyar University, Salem, India, PH-9994451525.
  E-mail:kaniezhil@yahoo.co.in
- Dr. C. Chandrasekar received his Ph. Ddegree from Periyar university. He is an Associate Professor of Computer Science, Periyar University, Salem, India. His areas of interest include Wireless networking, Mobile Computing, Computer Communications and Networks. He is a research guide at various universities in India. He has published more than 80 technical papers at various National & International conferences and 90 journals.
  E-mail:ccsekar@gmail.com

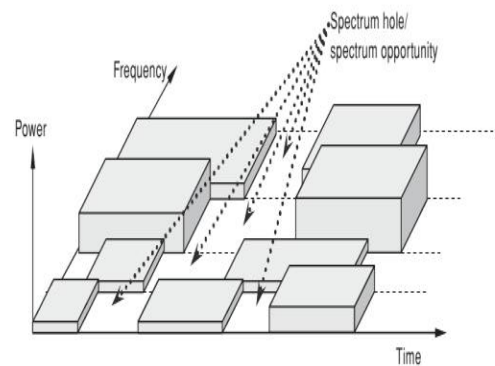

Fig. 1 The Spectrum Hole Concept

In addition large portion of licensed spectrum is underutilized. CR was created to solve this problem, by exploiting the existence of spectrum holes. Unlicensed users using CRs (Secondary Users), are aware of their spectrum environments and change their transmission and reception parameters to avoid interference with licensed spectrum users (Primary Users). This is shown in the Figure 1.

In addition to maximizing the efficiency of spectrum usage, CR's adaptation engine is supposed to improve wireless communication as a whole.

This paper presents a novel approach using Fuzzy logic system to utilize the available spectrum by the secondary users without interfering the primary user.

The paper is organized as follows; Section 2 and 3 defines cognitive radio and fuzzy logic system for its implementation. In section 4, opportunistic spectrum access by Fuzzy logic system to improve the spectrum efficiency. Section 5 and 6 presents the Knowledge Processing with Opportunistic Spectrum Access and simulation results. Finally, conclusions are presented in Section 7.

4## 2 COGNITIVE RADIO

The key enabling technology of dynamic spectrum access techniques is cognitive radio (CR) technology, which provides the capability to share the wireless channel with licensed users in an opportunistic manner. The term, cognitive radio, can formally be defined as follows :

A ''Cognitive Radio'' is a radio that can change its transmitter parameters based on interaction with the environment in which it operates [21]. From this definition, two main characteristics of the cognitive radio can be defined as follows [20]:

- Cognitive capability: It refers to the ability of the radio technology to capture or sense the information from its radio environment. Through this capability, the portions of the spectrum that are unused at a specific time or location can be identified. Consequently, the best spectrum and appropriate operating parameters can be selected.
  - Reconfigurability: The cognitive capability provides spectrum awareness whereas reconfigurability enables the radio to be dynamically programmed according to the radio environment.

CR networks are envisioned to provide high bandwidth to mobile users via heterogeneous wireless architectures and dynamic spectrum access techniques. This goal can be realized only through dynamic and efficient spectrum management techniques [8].

The main features of CR are listed as below [16,22]:
- Spectrum Sensing
- Spectrum Management
- Spectrum Mobility
- Spectrum Sharing

### 2.1 Spectrum Sensing

It detects the unused spectrum and sharing it without harmful interference with other users. It is an important requirement of the CR network to sense the spectrum holes. Primary users detection is found to be the most efficient way to detect the Spectrum holes [19].

### 2.2 Spectrum Management

It is the task of capturing the best available spectrum to meet the user requirements. CR should decide on the best spectrum band to meet the QoS requirements over all available spectrum bands, therefore spectrum management functions are required for CRs.

### 2.3 Spectrum Mobility

It is a process when the CR user exchanges its frequency of operation. CR networks target to use the spectrum in a dynamic manner by allowing the radio terminals to operate in the best available frequency band, maintaining seamless communication requirements during the transition to better spectrum.

### 2.4 Spectrum Sharing

It refers to providing the fair spectrum scheduling method, one of the major challenges in the open spectrum usage is the spectrum sharing. CRs have the capability to sense the surrounding environments and allow intended secondary user to increase QoS by opportunistically using the unutilized spectrum holes [19]. If a secondary user senses the available spectrum, it can use this spectrum after the primary licensed user vacates it.

## 3 FUZZY LOGIC SYSTEM

A fuzzy logic system (FLS) is unique in that it is able to simultaneously handle numerical data and linguistic knowledge. It is a nonlinear mapping of an input data (feature) vector into a scalar output, i.e., it maps numbers into numbers. Fuzzy set theory and fuzzy logic establish the specifics of the nonlinear mapping. It does this by starting with crisp set theory and dual logic and demonstrating how both can be extended to their fuzzy counterparts [1].

The fuzzy logic is one of the effective methods dealing with the partial state. In this sense, we are interested in that if we can apply fuzzy logic to differentiate the transmission states into different states which are followed with different membership degrees. With the concept of partial state in fuzzy logic, we can develop a fuzzy-based opportunistic spectrum access strategy to spectrum sharing. Therefore, the research problem is to apply a fuzzy-based optimal spectrum access strategy in the cognitive radio network.

Generally, Fuzzy Logic and Fuzzy decision making is divided into three consecutive phases namely Fuzzification, Fuzzy reasoning and Dufuzzification [17].

1. Fuzzification: The input variables are fuzzified using predefined membership functions (MF). Unlike in binary logic where only 0 and 1 are accepted, also numbers between 0 and 1 are used in fuzzy logic. This is accomplished with MFto which the input variables are compared. The output of fuzzification is a set of fuzzy numbers.
2. Fuzzy reasoning: Fuzzy numbers are fed into the predefined rule base that presents the relations of the input and output variables with IF – THEN Clauses. The output of the fuzzy reasoning is a fuzzy variable that is composed of the THEN clauses.
3. Dufuzzification: The output of the fuzzy reasoning is changed into a non-fuzzy number that represents the actual output of the system.

Fuzzy sets theory is an excellent mathematical tool to handle the uncertainty arising due to vagueness [3,15]. Figure 2 shows the structure of a fuzzy logic system.

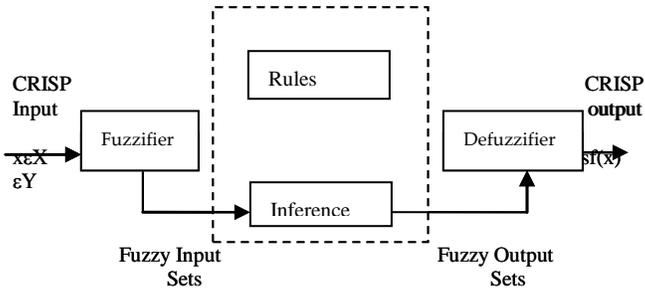

Fig. 2 The Structure of a Fuzzy Logic System

Since there is a need to "fuzzify" the fuzzy results we generate through a fuzzy system analysis i.e., we may eventually find a need to convert the fuzzy results to crisp results. Here, we may want to transform a fuzzy partition or pattern into a crisp partition or pattern; in control we may want to give a single-valued input instead of a fuzzy input command. The "dufuzzification" has the result of reducing a fuzzy set to a crisp single-valued quantity, or to a crisp set.

Consider a fuzzy logic system with a rule base of M rules, and let the lth rule be denoted by $R_l$. Let each rule have p antecedents and one consequent (as is well known, a rule with q consequents can be decomposed into rules, each having the same antecedents and one different consequent), i.e., it is of the general form [18]

$$R_l : \text{IF } u_1 \text{ is } F_l^1 \text{ and } u_2 \text{ is } F_l^2 \text{ and } \ldots \text{ and } u_p \text{ is } F_l^p, \text{ THEN } v \text{ is } G^l.$$

where $u_k, K=1,\ldots p$ and $v$ are the input and output linguistic variables, respectively. Each $F_k^l$ and $G^l$ are subsets of possibly different universes of discourse. Let $F_k^l \subset U_k$ and $G^l \subset V$. Each rule can be viewed as a fuzzy relation $R_l$ from U to a set V where U is the Cartesian product $U = U_1,\ldots,U_p$. $R_l$ itself is a subset of the Cartesian product U X V is $\{(x,y) : x \in U, y \in V\}$, where $x \equiv (x_1, x_2, \ldots, x_p)$ and $x_k$ and y are the points in the universes of discourse $U_k$ and V of $u_k$ and v.

While applying a singleton fuzzification, when an input $X = \{x_1, x_2, x_3, \ldots x_p\}$ [11] is applied, the degree of firing corresponding to the lth rule is given by

$$x^* = \frac{\sum_{i=1}^{n} x_i \cdot \mu(x_i)}{\sum_{i=1}^{n} \mu(x_i)} \quad (1)$$

Where * denotes a T-norm, n represents the number of elements, $x_i$'s are the elements and $\mu(x_i)$ is its membership function [2,3]. There are many kinds of dufuzzification methods, but we have chosen the centre of sets method for illustrative purpose. It computes a crisp output for the FLS by first computing the centroid, of every consequent set $G^l$ and, then computing weighted average of these centroids. The weight corresponding to the $l$th rule consequent centroid [4] is the degree of firing associated with the lth rule, $T_{i=1}^{p} \mu_{F_l^l}(x_1)$ so that

$$y_{cos}(x') = \frac{\sum_{l=1}^{M} C_{G^l} T_{i=1}^{p} \mu_{F_l^l}(x_1')}{\sum_{l=1}^{M} T_{i=1}^{p} \mu_{F_l^l}(x_1')} \quad (2)$$

where M is the number of rules in the FLS.

## 4 OPPORTUNISTIC SPECTRUM ACCESS USING FUZZY LOGIC

We design the fuzzy logic for opportunistic spectrum access using cognitive radio. In this paper, we are selecting the best suitable secondary users to access the available users without any interference with the primary users. This is collected based on the following three antecedents i.e., descriptors. They are
Antecedent 1: Spectrum Utilization Efficiency
Antecedent 2: Degree of Mobility
Antecedent 3: Distance of Secondary user to the
            Primary user.

Fuzzy logic is used because it is a multi-valued logic and many input parameters can be considered to take the decision. Generally, the secondary user with the furthest distance to the primary user or the secondary user with maximum spectrum utilization efficiency can be chosen to access spectrum under the constraint that no interference is created for the primary user [14]. In our approach, we combine the three antecedents to allocate spectrum opportunistically inorder to find out the optimal solutions using the fuzzy logic system.

Mobility of the secondary user plays a vital role in the proposed work. Wireless systems also differ in the amount of mobility that they have to allow for the users. The ability to move around while communicating is one of the main charms of wireless communication for the user. Spectrum Mobility is defined as the process when a cognitive radio user exchanges its frequency of operation. The movement of the secondary user leads to a shift of the received frequency, called the Doppler shift. When the secondary user is moving at a velocity v m/s, it causes the Doppler effect.

The Doppler effect leads to a shift of the received frequency $f_D$ by the amount $v$, so that the received frequency is given by:

$$f_D = \frac{v \cos \theta}{c} f_c \text{ is } f_c - v \quad (3)$$

where $f_D$ is the Doppler shift, $\theta$ denotes the angle between the velocity vector v of the MS and the direction of the wave at the location of the MS, c is the wave velocity, and f is carrier frequency [6]. Obviously, the frequency shift depends on the direction of the wave, and must lie in the range $f_c - \ldots f_c + v_{max}$, where $v_{max}$ is $f_c v/c$.

Detecting signal from the primary user can be reduced using the Mobility. The secondary user should detect the primary signal which determines the spectrum that is unused. If the secondary user fails to detect the primary signal, then it will not determine exactly the spectrum that is unused. Thereby it leads to interference to the adjacent users. This is referred as the hidden node problem.

Besides, we consider the distance between the primary user and the secondary users. We consider the locally measured SNR as a proxy for distance, it is convenient to represent $r_p, r_n, r_{dec}$ in terms of the SNR in dB measured at those points. Actually we do not know the location of the primary user. Therefore we must specify who is measuring the SNR at each distance [6]. We consider $\gamma_{dec}, \gamma_p$ to be measured as a primary receiver and $\gamma_n$ as a secondary transmitter. We define :

$$\gamma_n = 10 \log\left(\frac{P_1 g(R)}{\sigma_l^2}\right) \quad (4)$$

where $P_1$ is the transmit power of the primary user and $\sigma_l^2$ is the noise power measured at the secondary user. From the equation (4), we can derive the distance R between the primary user and the secondary user.

Distance between the primary user and the secondary users Were calculated using the formula

$$D_i = \frac{d_i}{max_{i=1}^{20}\{d_i\}} \quad (5)$$

$$d_i = \sqrt{(x_i - x_p)^2 + (y_i - y_p)^2} \quad (6)$$

where $x_p$ and $y_p$ are the distances of the primary users and $x_i$ and $y_i$ are the distances of the secondary users from the primary user.

We apply different available spectrum inorder to find out the spectrum efficiency which is the main purpose of the opportunistic spectrum access strategy. Hence, we calculate the spectrum efficiency $\eta_s$ as the ratio of average busy spectrum over total available spectrum owned by secondary users, i.e.,

$$\eta_s = \frac{n_{busy\_spectrum}}{n_{ava\_spectrum}} \quad (7)$$

where $n_{busy\_spectrum}$ is the number of busy spectrum used at time t for secondary user and $n_{ava\_spectrum}$ is the total available spectrum respectively [7].

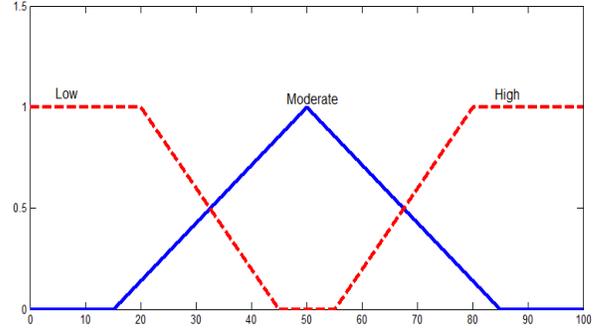

(3. a) Membership function (MF) used to represent the antecedent 1

The linguistic variables are used to represent the spectrum utilization efficiency, distance and degree of mobility are divided into three levels: low, moderate, and high. while we use 3 levels, i.e., near, moderate, and far to represent the distance.

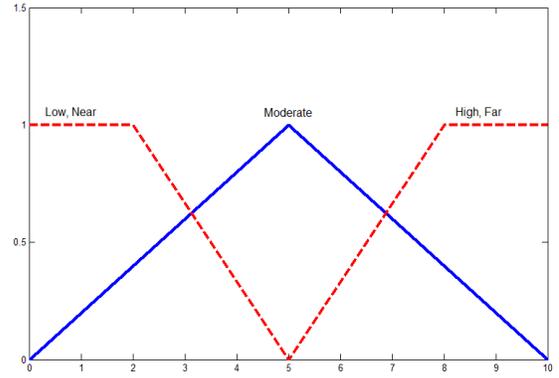

(3.b) Membership function (MF) used to represent the antecedent 2

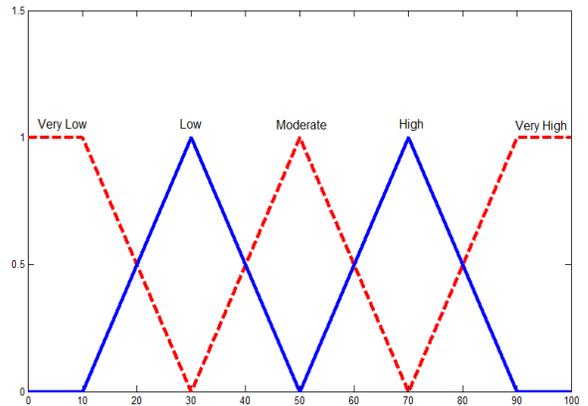

(3. c) Membership function (MF) used to represent the antecedent 3

The consequence, i.e., the possibility that the secondary user is chosen to access the spectrum is divided into five levels which are very low, low, medium, high and very high [15]. We use trapezoidal membership functions (MFs) to represent near, low, far, high, very low and very high, and triangle MFs to represent moderate, low, medium and high. MFs are shown in Fig. 3a, 3b,

3c. Since we have 3 antecedents and fuzzy subsets, we need setup $3^3 = 27$ rules for this FLS.

## 5 KNOWLEDGE PROCESSING WITH OPPORTUNISTIC SPECTRUM ACCESS

The proposed Fuzzy Logic System (FLS) takes decision based on the key parameters according to the predefined rules i.e., the three antecedents and its consequence as shown in the Table 1.

Table 1.
Fuzzy Rules

1. If (Antecedent1 is Low && Antecedent2 is Low && Antecedent3 is Near) then Consequence is Very Low.
2. If (Antecedent1 is Low && Antecedent2 is Low && Antecedent3 is Moderate) then Consequence is Low.
3. If (Antecedent1 is Low && Antecedent2 is Low && Antecedent3 is Far) then Consequence is Low.
4. If (Antecedent1 is Low && Antecedent2 is Moderate && Antecedent3 is Near) then Consequence is Very Low.
5. If (Antecedent1 is Low && Antecedent2 is Moderate && Antecedent3 is Moderate) then Consequence is Low.
6. If (Antecedent1 is Low && Antecedent2 is Moderate && Antecedent3 is Far) then Consequence is Medium.
7. If (Antecedent1 is Low && Antecedent2 is High && Antecedent3 is Near) then Consequence is Very Low.
8. If (Antecedent1 is Low && Antecedent2 is High && Antecedent3isModerate) then Consequence is Low.
9. If (Antecedent1 is Low && Antecedent2 is High && Antecedent3 is Far) then Consequence is Medium.
10. If (Antecedent1 is Moderate && Antecedent2 is Low && Antecedent3 is Near) then Consequence is Very Low.
11. If (Antecedent1 is Moderate && Antecedent2 is Low && Antecedent3 is Moderate) then Consequence is Medium.
12. If (Antecedent1 is Moderate && Antecedent2 is Low && Antecedent3 is Far) then Consequence is High.
13. If (Antecedent1 is Moderate && Antecedent2 is Moderate && Antecedent3 is Near) then Consequence is Very Low.
14. If (Antecedent1 is Moderate && Antecedent2 is Moderate && Antecedent3 is Moderate) then Consequence is Medium.
15. If (Antecedent1 is Moderate && Antecedent2 is Moderate && Antecedent3 is Far) then Consequence is High.
16. If (Antecedent1 is Moderate && Antecedent2 is High && Antecedent3 is Near) then Consequence is Very Low.
17. If (Antecedent1 is Moderate && Antecedent2 is High && Antecedent3 is Moderate) then Consequence is Low.
18. If (Antecedent1 is Moderate && Antecedent2 is High && Antecedent3 is Far) then Consequence is High.
19. If (Antecedent1 is High && Antecedent2 is Low && Antecedent3 is Near) then Consequence is Low.
20. If (Antecedent1 is High && Antecedent2 is Low && Antecedent3 is Moderate) then Consequence is High.
21. If (Antecedent1 is High && Antecedent2 is Low && Antecedent3 is Far) then Consequence is Very High.
22. If (Antecedent1 is High && Antecedent2 is Moderate && Antecedent3 is Near) then Consequence is Low.
23. If (Antecedent1 is High && Antecedent2 is Moderate && Antecedent3 is Moderate) then Consequence is High.
24. If (Antecedent1 is High && Antecedent2 is Moderate && Antecedent3 is Far) then Consequence is Very High.
25. If (Antecedent1 is High && Antecedent2 is High && Antecedent3 is Near) then Consequence is Very Low.
26. If (Antecedent1 is High && Antecedent2 is High && Antecedent3 is Moderate) then Consequence is High.
27. If (Antecedent1 is High && Antecedent2 is High && Antecedent3 is Far) then Consequence is High.

Since we chose a single consequent for each rule to form a rule base, we averaged the centroids of all the responses for each rule and used this average in place of the rule consequent centroid. Doing this leads to rules that have the following form:

$R'$: If Degree of mobility ($x_1$) is $F_l^1$; and its distance between primary user and the secondary users ($x_2$) is $F_l^2$; and the spectrum utilization efficiency of the secondary user ($x_3$) is $F_l^3$, then the Possibility (y) choosing the available spectrum is $c_{avg}^l$, where l is 1,2,..27 and $c_{avg}^l$ is defined as follows:

$$c_{avg}^l = \frac{\sum_{i=1}^{5} w_i^l c^i}{\sum_{i=1}^{5} w_i^l} \qquad (8)$$

in which $w_i^l$ is the number of choosing linguistic label i for the consequence of rule l and $c^i$ is the centroid of the ith consequence set (i: 1; 2; ...; 5; l: 1; 2; ...; 27). Table 2 provides $c^i$ for each rule. For every input ($x_1, x_2, x_3$), the output y($x_1, x_2, x_3$) of the designed FLS [5] is computed as

$$y(x_1, x_2, x_3) = \frac{\sum_{i=1}^{27} \mu_{F_1^l(x_1)} \mu_{F_2^l(x_2)} \mu_{F_3^l(x_3)} c_{avg}^l}{\sum_{i=1}^{27} \mu_{F_1^l(x_1)} \mu_{F_2^l(x_2)} \mu_{F_3^l(x_3)}} \quad (9)$$

which gives the possibility that a secondary user is selected to access the available spectrum. By using (9), the secondary user with the highest possibility would be chosen to access the available spectrum.

Table. 2
$c_{avg}^l$ to each corresponding rule

| Rule# | $c_{avg}^l$ |
|---|---|
| 1 | 28.59 |
| 2 | 25.90 |
| 3 | 24.23 |
| 4 | 22.43 |
| 5 | 22.98 |
| 6 | 24.68 |
| 7 | 16.95 |
| 8 | 19.70 |
| 9 | 22.06 |
| 10 | 43.08 |
| 11 | 40.20 |
| 12 | 38.98 |
| 13 | 40.89 |
| 14 | 38.47 |
| 15 | 39.16 |
| 16 | 36.50 |
| 17 | 34.15 |
| 18 | 40.26 |
| 19 | 58.62 |
| 20 | 55.12 |
| 21 | 54.75 |
| 22 | 56.99 |
| 23 | 53.81 |
| 24 | 53.92 |
| 25 | 54.05 |
| 26 | 53.72 |
| 27 | 52.12 |

The weighted average value for each rule is given in the table 2. Table 3 gives the three Descriptors and Possibility for four Secondary users. As listed in Table 3, at a particular time, values of three descriptors and possibility for four secondary user i.e., the secondary user chosen to the access the available spectrum is (SU4), the secondary user with the highest spectrum utilization is (SU2), the secondary user having the furthest distance to the primary user is (SU3), and the secondary user with the lowest mobility is (SU4).

From table 3, we see that SU1 has 61%, SU2 has 90% of Spectrum utilization efficiency, SU3 only achieves 70.89% and SU4 achieves 83% of spectrum Utilization. Similarly, if we consider the Distance of primary user to the secondary users, SU1 has 8.01, SU2 has 2.16, SU3 has the furthest distance from the primary user 12.80and SU4 has 6.02.

Table 3.
Three Descriptors and Possibility for Four Secondary users

| Parameters | SU1 | SU2 | SU3 | SU4 |
|---|---|---|---|---|
| Distance from Primary user to secondary users | 8.01 | 2.16 | 12.80 | 6.02 |
| Possibility | 28.59 | 19.70 | 40.89 | 58.62 |
| Degree of Mobility | 6.6667 | 9.2528 | 5.2229 | 1.2420 |
| Spectrum Utilization Efficiency | 61.0014 | 90.4191 | 70.8922 | 83.0274 |

Even though, if SU3 has furthest distance from the primary user but it has lowest spectrum utilization (70.89%) and lowest degree of mobility. Thus, the secondary user will select the spectrum for accessing based on the highest possibility rather than the highest spectrum utilization and the furthest distance from the primary user.

## 6 SIMULATION RESULTS

In this section, we present simulation results on the performance of our proposed work based on Fuzzy logic System and proposed sensing framework. In the proposed work, we are choosing the available channel with the high possibility and high spectrum utilization efficiency.

To validate our approach, we randomly generated 20 secondary users over an area of 100 X 100 meters. The primary user was placed randomly in this area. Three descriptors were randomly generated for each secondary user. More specifically, the spectrum utilization efficiency of each secondary user was a random value in the interval [0,100] and its mobility degree in [0,10]. Distances to the primary users were normalized to [0,10].

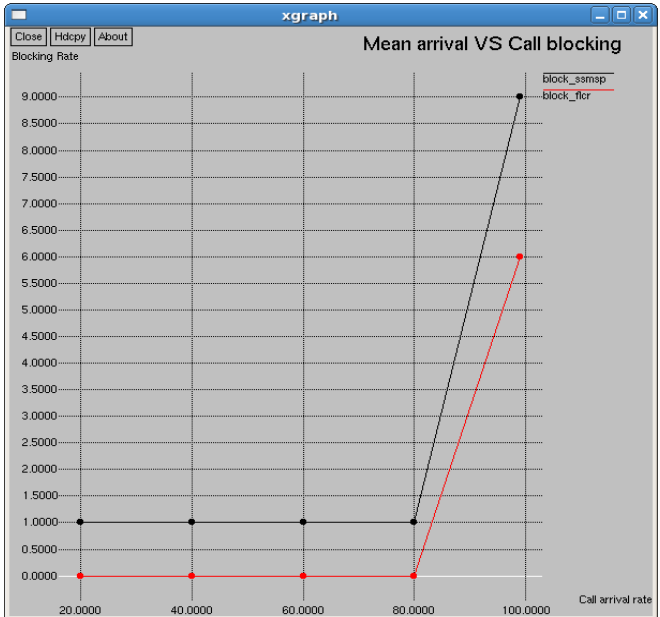
Fig. 4  Comparison of Mean Arrival VS Call blocking using normal spectrum utilization method and Fuzzy Logic System

Figure 4 shows the comparison of call blocking of the service provider using the Fuzzy logic system and normal Spectrum Utilization (NSU) [7-12]. As the call arrival rate increases the blocking rate gets decreased. Traffic rate increases along with the call blocking rate. Here, the Fuzzy Logic System shows less call blocking rate when compared to the normal Spectrum utilization approach.

Using the fuzzy logic system, the free spectrum is calculated in order to find out the available spectrum which should be utilized based on the service provider's request. When compared to the normal spectral access approach, the Fuzzy Logic System provide more availability of the spectrum. The Figure 5 demonstrates the accuracy of the available spectrum.

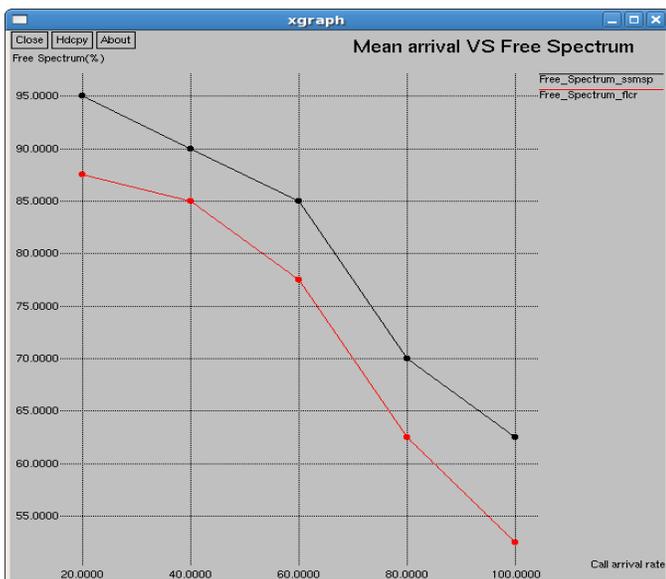
Fig. 5  Comparison of Mean Arrival VS Free Spectrum using normal spectrum utilization and Fuzzy Logic

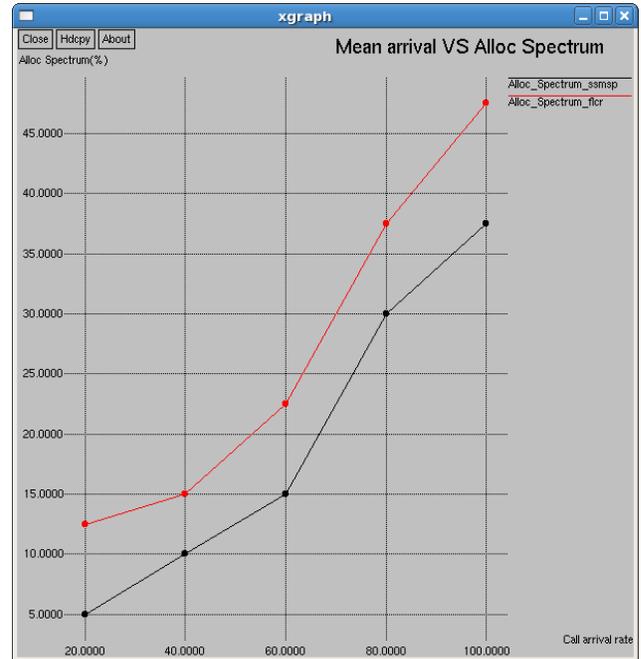
Fig. 6 Comparison of Mean Arrival VS Allocated Spectrum using normal spectrum utilization and Fuzzy Logic System

As illustrated in the Figure 6, Spectrum is allocated based on the free spectrum information. Using the fuzzy logic, spectrum is allocated on the opportunistic manner. Based on the Free Spectrum calculation, spectrum is allocated efficiently using Fuzzy Logic System rather than the normal spectrum utilization approach.

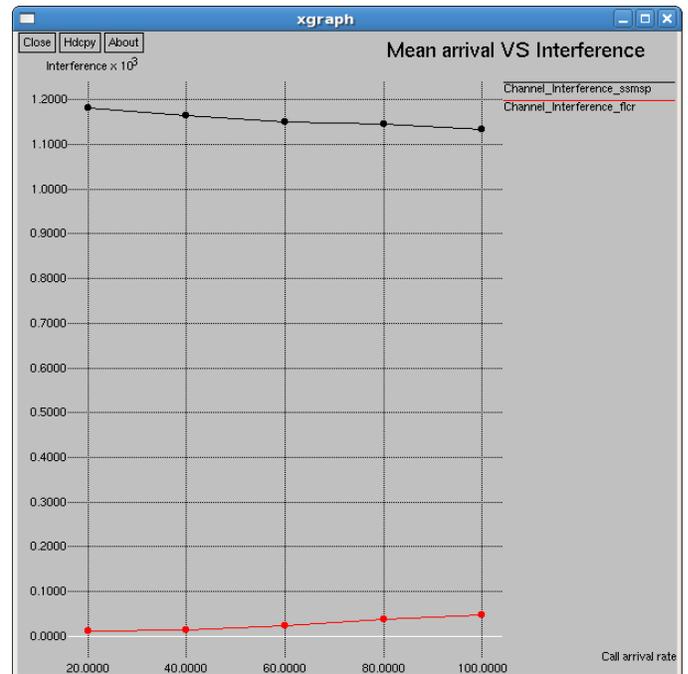
Fig.7  Comparison of Mean Arrival VS Interference using normal spectrum utilization and Fuzzy Logic System

Interference is the key factor that limits the performance of wireless networks. Spectrum managers are concerned with managing interference and in establishing the methods, techniques, information and processes needed to protect users and uses from harmful interference. Harmful interference arises in radio systems when a transmitter's ability to communicate with its intended receiver(s) is limited because of the transmissions of other transmitters. Interference is calculated using the formula :

$$interference = |frmax - frmin|$$

As illustrated in the Figure 7, interference gets decreased as the call arrival rate increases. When compared to the normal spectral utilization approach, the Fuzzy Logic System provides the minimum interference occurrence.

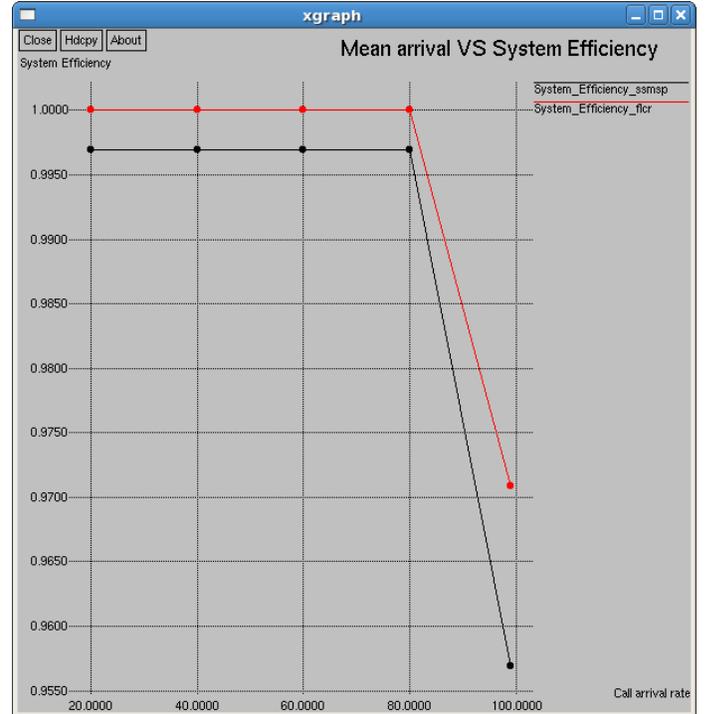

Fig. 9 Comparison of Mean arrival VS System Efficiency using normal spectrum utilization and Fuzzy Logic System

The system efficiency decreases when the traffic rate is beyond the system capacity. There is an increase in the system efficiency using the Fuzzy logic system. When compared to the NSU, the fuzzy logic system gives the better system efficiency. This fluctuation in the system efficiency is shown in the figure 9.

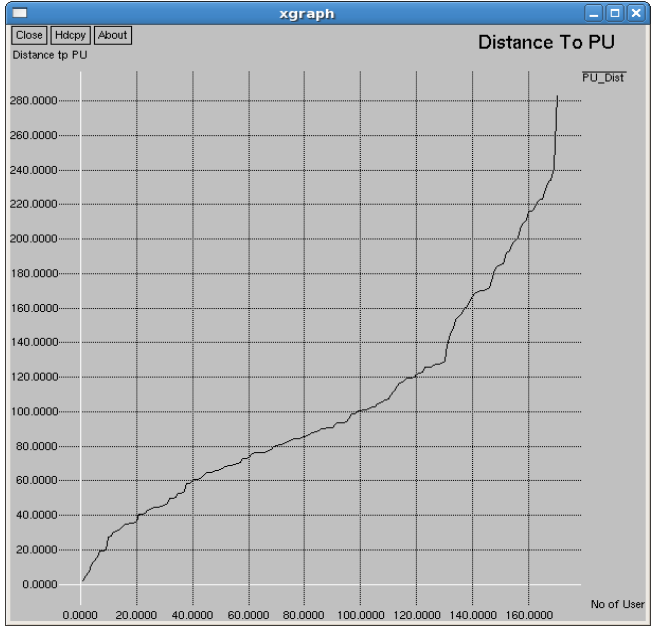

Fig. 8 Distance from Primary user to the Secondary users using Fuzzy Logic System

Distance from primary user to the secondary user is calculated in order to find out the maximum distance between these users. So that, secondary users which have the furthest distance will have higher chance to access the spectrum. This is shown using the Figure 8. Similarly, the chance is getting increased when required spectrum is low compared to the available spectrum.

The System efficiency $\eta_{sys}^{(i)}$ is defined as Probability efficiency metric for service provider is determined by the processed traffic intensity and the total traffic loaded to service provider within the observation time. Thus, $\eta_{sys}^{(i)}$ is calculated by

$$\eta_{sys}^{(i)} = \frac{E_p^{(i)}}{E_{in}^{(i)}}$$

Where $E_p^{(i)}$ is the processed traffic intensity in Erlang for service provider i and $E_{in}^{(i)}$ is the total traffic loaded to the service provider i within the observation time t.

Maximum channel utilization is carried out in the proposed work. From the above results, we can confirm that spectrum access decision is tradeoffs among three descriptors chosen to design the FLS. Therefore, the secondary user with the highest spectrum utilization or the secondary user furthest from the primary user is guaranteed to access the spectrum.

Since we use the distributed spectrum sharing architecture, a distributed entity such as base stations in cognitive wireless networks collects information about three descriptors and available spectrum bands from secondary users and builds a spectrum map. With this, our designed FLS is used to control the spectrum assignment and access procedures in order to prevent multiple users from colliding in overlapping spectrum portions.

The Spectrum Efficiency (Channel tilization) $\eta_s^{n_{(sp)}}$ is defined as the ratio of average busy channels over total channels owned by service providers. It corresponds to

$$\eta_s^{n_{(sp)}} = \lim \frac{1}{t} \int_0^t \frac{n_{busy}^{n_{(sp)}}(t)}{N_{ch-total}^{n_{(sp)}}(t)} dt$$

where $n_{busy}^{n_{(sp)}}(t)$ is the number of channels used at time t for service provider $n_{(sp)}$ and $N_{ch-total}^{n_{(sp)}}(t)$ is the total number of total channels owned by service provider $n_{(sp)}$.

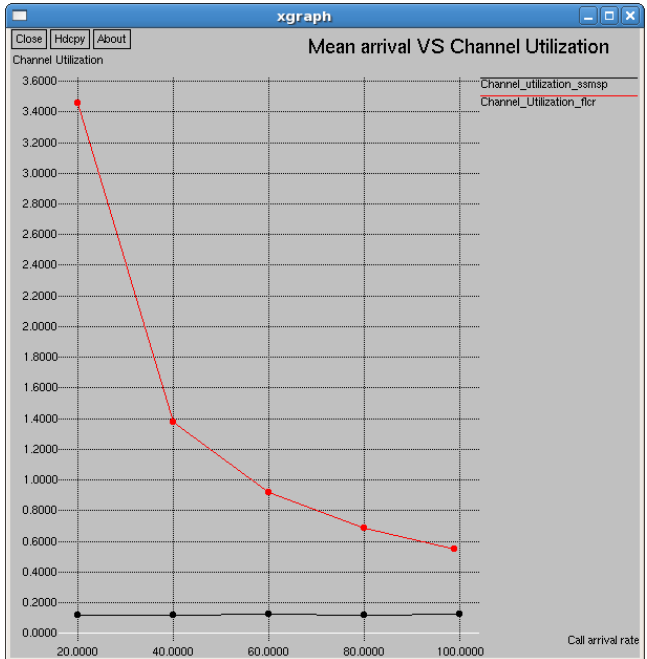

Fig. 10 Comparison of Mean arrival VS Channel utilization using normal spectrum utilization and Fuzzy Logic System

Higher Spectrum efficiency is estimated because the call blocking rate is lower; thus more calls can contribute to the spectrum utilization. Figure 10 clearly shows that the fuzzy logic system provides the better spectrum utilization compared to the normal spectrum utilization approach.

## 7 CONCLUSION

A Fuzzy logic system approach is proposed to control the opportunistic spectrum access for secondary users in cognitive radio networks. The secondary user is selected for accessing the available spectrum is based on three antecedents namely spectrum utilization efficiency of the secondary user, degree of mobility and distance from primary user to the secondary users.

We are trying to enhance the spectrum aware communication by this work in order to fulfill the present status of the spectrum utilization and avoiding the spectrum scarcity. Spectrum allocation method by opportunistic spectrum access scenario using cognitive radio was analyzed and simulated using NS2 Simulator in order to validate our approach.

Our approach provides a way for secondary users to get as much as spectrum bands for access with the help of the distance from Primary user to the secondary users. It also shows that efficient spectrum utilization by solving the spectrum mobility problem which provides a high QoS of cognitive radio systems using the Fuzzy logic System rather than the normal Spectral Utilization approach.

Hence, the secondary users can sense and utilize the unutilized spectrum of the primary user. In this work, a fuzzy logic based system provides a way for the secondary user can opportunistically use the spectrum efficiently and thus avoids spectrum scarcity.

## ACKNOWLEDGMENT

The First Author extends her gratitude to UGC as this work was supported in part by Basic Scientist Research (BSR) Non SAP Scheme, under grant reference no.11-142/2008(BSR) UGC XI Plan.